\documentclass{cimento}

\usepackage{graphicx}  

\usepackage{siunitx}
\usepackage[version=4]{mhchem}

\title{Processing of non-constant baseline pulses: a matrix technique}
\author{C.~Ferrari\from{z}\from{t}\thanks{cecilia.ferrari@gssi.it}, 
M.~Borghesi\from{ins:x}\from{ins:y},
M.~Faverzani\from{ins:x}\from{ins:y},
E.~Ferri\from{ins:x}\from{ins:y},
A.~Giachero\from{ins:x}\from{ins:y} \atque A.~Nucciotti\from{ins:x}\from{ins:y}}

\instlist{
	\inst{z} Gran Sasso Science Institute (GSSI), I-67100 L’Aquila, Italy
	\inst{t} INFN - Laboratori Nazionali del Gran Sasso, Assergi (L’Aquila) I-67010 - Italy
	\inst{ins:x} Dipartimento di Fisica ``G. Occhialini”, Universit\`a di Milano - Bicocca, Milan 20126, Italy\inst{ins:y} Istituto Nazionale di Fisica Nucleare (INFN), Sezione di Milano-Bicocca, Milan 20126, Italy}

\preprinttrue
\shortauthor{C. Ferrari \etal}

\begin{document}
\sloppy

\maketitle

\begin{abstract}
	For a high source activity experiment, such as HOLMES, non-constant baseline pulses could constitute a great fraction of the data-set.  
	We test the optimal filter matrix technique, proposed to process these pulses, on simulated responses of HOLMES microcalorimeters
\end{abstract}

\section{The HOLMES experiment}

The HOLMES experiment aims at measuring the electron neutrino mass by studying the spectrum of the \ce{^{163}Ho} EC decay. A non null neutrino mass, indeed, modifies the part of the spectrum at higher energies in its shape and end-point value. In order to reach \SI{2}{\eV} of sensitivity, the HOLMES experiment will exploit $1024$ microcalorimeters each composed by a \ce{Mo/Cu} Transition-Edge Sensor (TES) and a gold absorber implanted with \SI{300}{Bq} of \ce{^{163}Ho}~\cite{ref:ferri}.
Whenever a \ce{^{163}Ho} decay occurs, the energy released in the absorber is measured by the TES as a temperature variation and converted into a current pulse.
By evaluating these pulses amplitudes, it is possible to perform the \ce{^{163}Ho} spectrum measurement.

\section{Pulse fitting procedure: the optimal filter matrix}

The simplest approach that can be employed for the evaluation of the pulses amplitudes is to fit the triggered data with a model.
The implementation of this fitting technique, briefly explained in this section, will lead to the construction of the matrix filter described in~\cite{ref:fowler}. 

Suppose that \emph{\bf d} is the vector representing the sampled pulse and that \emph{\bf m}(\emph{\bf p}) is its model, depending on the parameters vector \emph{\bf p}. Presume moreover that each sample of {\bf d} is affected by Gaussian fluctuation around its value.
Since the entries of the {\bf d} vector are not independent from each other, in order to write the correct Likelihood for the fit, one has to take into consideration the covariance matrix $\emph{R}$. It is easy to prove~\cite{ref:mez} that the Likelihood results to be:
\begin{equation}
L \propto \exp{\left[-({\bf d}-{\bf m})^{\text{T}}R^{-1}({\bf d}-{\bf m})\right]}.
\label{eq:like}
\end{equation}
Assuming that the only feature changing with the energy is the pulse amplitude, the model {\bf m} can be rewritten as:
\begin{equation}
{\bf m} = M{\bf p}^{\text{T}},
\label{eq:M}
\end{equation}
where $M$ is the matrix collecting the model components. For example, let's assume we want to fit a pulse raising on a constant baseline. In this case, the matrix $M$ would be composed of two columns: the pulse model {\bf s} with unitary amplitude and a vector of ones modeling the flat baseline. Therefore, the model {\bf m} would result in:
$$
{\bf m} = p_{1}\times{\bf s} + p_{2}\times(1,\dots,1), 
$$
identifying $p_{1}$ as the amplitude value and $p_{2}$ with the baseline level.

In order to maximize the Likelihood in~(\ref{eq:like}), we minimize the quantity $Q^{2} = ({\bf d}-M{\bf p})^{\text{T}}R^{-1}({\bf d}-M{\bf p})$ in respect of the parameters vector {\bf p}. This results in
\begin{equation}
{\bf p}^{\text{T}} = (M^{\text{T}}R^{-1}M)^{-1}(M^{\text{T}}R^{-1}) {\bf d},
\end{equation}
where the inverse of $M^{\text{T}}R^{-1}M$ always exists since it is a positive definite symmetric matrix. Therefore, we can define the filter matrix $q := (M^{\text{T}}R^{-1}M)^{-1}(M^{\text{T}}R^{-1})$ where $R$ is the Toeplitz noise covariance matrix computed with an enough wide sample of noise records and $M$ is the collection of columns properly modeling the processed data.

\section{Matrix filter application to TES pulses}

\begin{figure}
\centering
\includegraphics[scale=0.41]{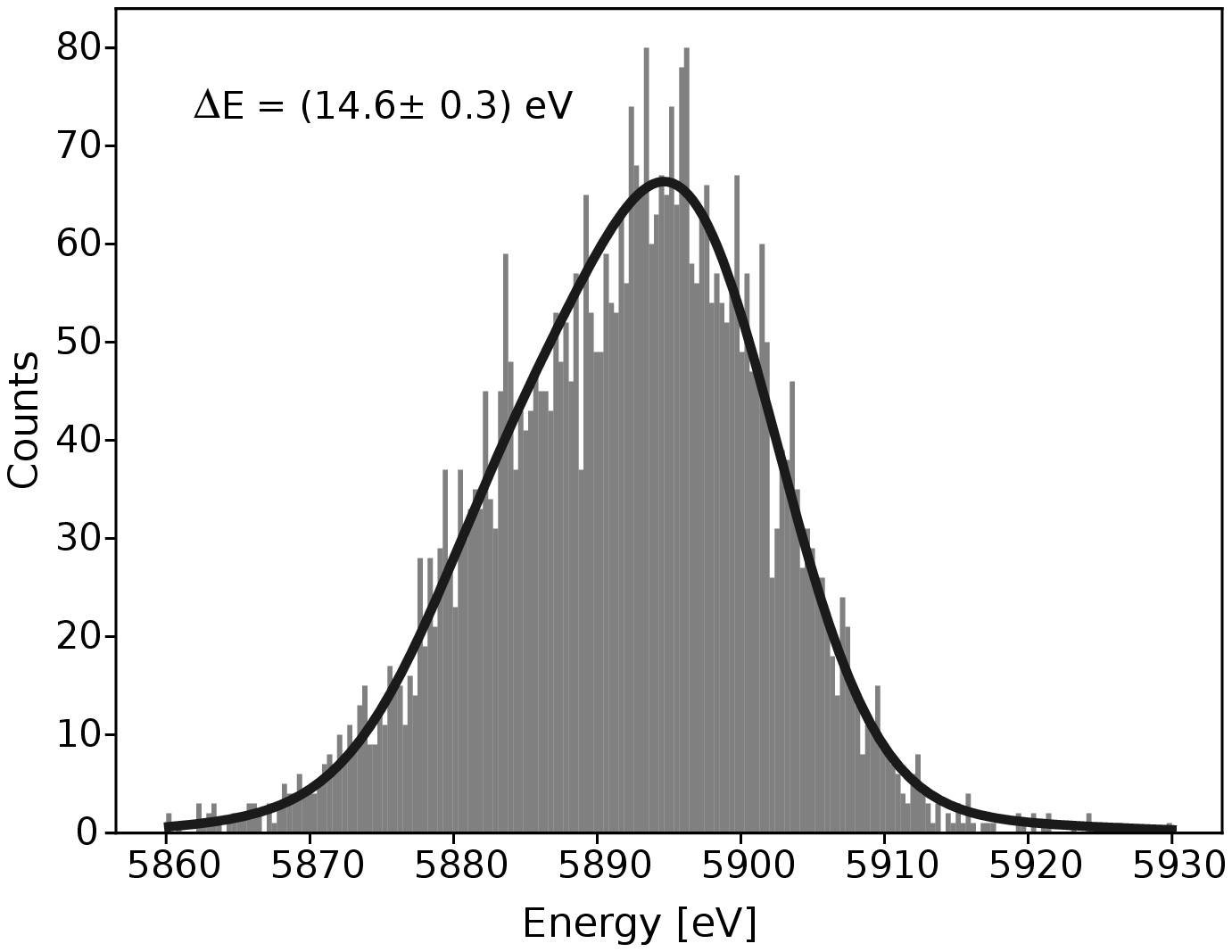}
\includegraphics[scale=0.41]{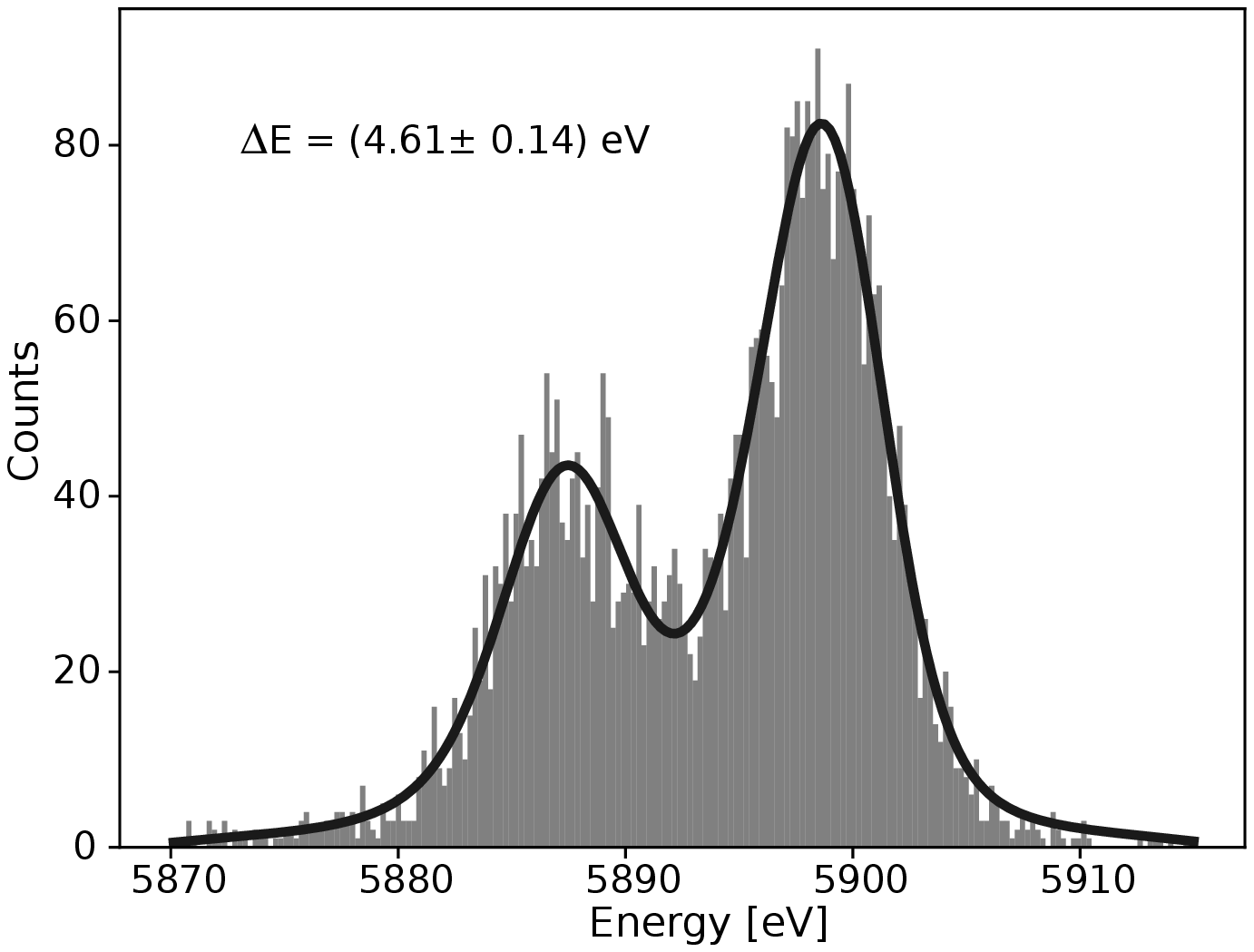}
\caption{\label{fig:fwhm} \ce{^{55}Mn} peak measured by estimating the energies from raw TESs pulses (left) and by applying the optimal filter matrix technique (right).}
\end{figure}

We have analyzed with the matrix filtering technique the pulses of HOLMES TESs probed with a source of \ce{^{55}Fe}, considering only the records with single signals raising on a constant baseline. Therefore, the matrix $M$  of equation~(\ref{eq:M}) was implemented with only two columns: the first one modeling the signal shape with unitary amplitude and the second one representing the flat baseline. The resulting energy spectrum, compared to that obtained with a rough estimation of the pulses amplitude, is shown in figure~(\ref{fig:fwhm}) where the \ce{^{55}Mn} peak has been fit with a combination of Lorentzian functions.
The obtained energy resolution (\SI{4.61}{\eV} at about \SI{6}{\keV}) is compatible with that found with the most common pulse processing technique (the optimal filter~\cite{ref:gatti}) as shown in~\cite{ref:marco}.

\section{Exponential constrain for non-constant baseline pulses}

\begin{figure}
\centering
\includegraphics[scale=0.41]{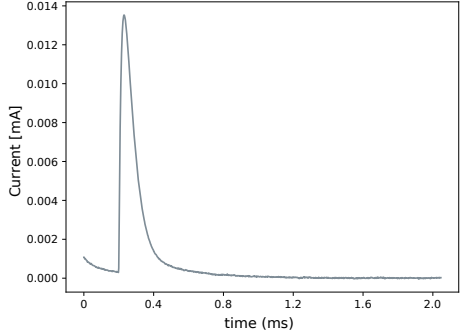}
\caption{\label{fig:example} Simulated HOLMES TES pulse raising on a previous signal tail.}
\end{figure}

Since \SI{300}{Bq} for each detector is a high source activity, the probability of triggering pulses rising on the tail of prior signals is non-zero. In order to probe if the matrix technique can correctly process also this kind of pulses, we performed a simulation of the detector response, considering the HOLMES TES physical parameters enlisted in~\cite{ref:marco}.
Moreover, in order to test different contribution of the tails over which the studied pulses raised, we varied the arrival time of the prior signal. An example of these simulated pulses is reported in figure~(\ref{fig:example}).

\begin{figure}
\includegraphics[width=.95\textwidth]{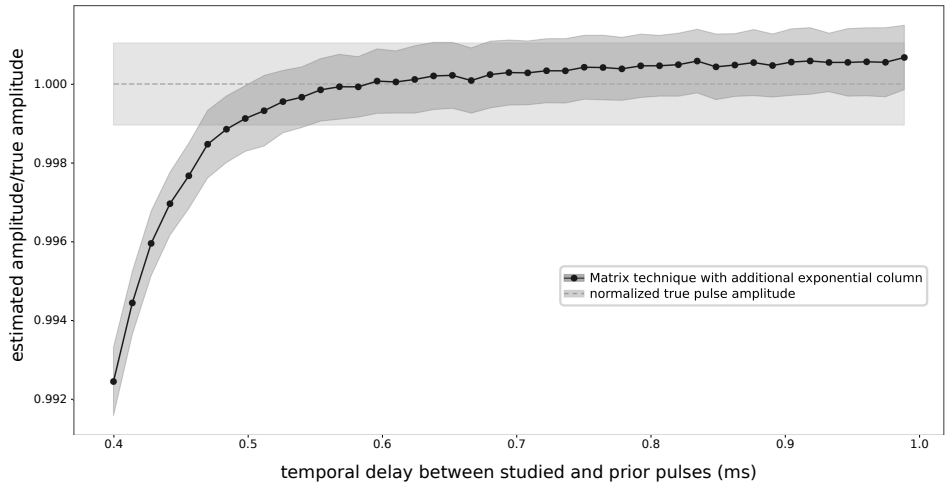}
\caption{\label{fig:grafico} Results of matrix filtering technique application to simulated non-constant baseline pulses (in dark gray with its $1\sigma$ band) according to the temporal difference between the two signals. As a comparison, the evaluated mean amplitude of pulses rising on constant baseline is reported (in light gray with its $1\sigma$ band).}
\end{figure}

In this case the $M$ matrix included one more column in respect of that exploited for the flat baseline pulses. This additional element was introduced to model the prior signal tail that in the sake of simplicity was assumed as a simple exponential decay. The characteristic temporal decay constant was set to \SI{250}{\micro\s}, which corresponds to the one measured in HOLMES TES pulses~\cite{ref:marco}.

Figure~(\ref{fig:grafico}) shows the results of this study. 
For quite every simulated temporal delay the matrix technique with the additional exponential constraint evaluates the pulses amplitude quite correctly.
As we expected, at smaller delay times, where the contribution of prior pulse tail is more important, the estimated pulse amplitude deviates from its true value.
Moreover the obtained standard deviation (which is reported in figure~(\ref{fig:grafico}) with the bands) is compatible to that found in processing pulses rising on a constant baseline.

\section{Conclusions}

In conclusion, the matrix filtering technique is a powerful tool for microcalorimeter pulse processing.  In fact, it provides very good energy resolutions and it also allows the correct amplitude evaluation of the non-constant baseline events.
A good method to optimize the non-constant pulses amplitude estimation is that of fitting for each record the baseline region preceding the signal to better choose the exponential decay constant. This obviously implies a greater computational cost. For this reason further studies on the optimization of the exponential decay constant are required.

\acknowledgments
This work was supported by the European Research Council (FP7/2007-2013), under Grant Agreement HOLMES no.340321, and by the INFN Astroparticle Physics Commission 2 (CSN2). We also acknowledge the support from the NIST Innovations in Measurement Science program for the TES detector development.

\end{document}